\documentclass[10pt,conference]{IEEEtran}

\usepackage{graphicx}% Include figure files
\usepackage{dcolumn}% Align table columns on decimal point
\usepackage{bm}% bold math
\usepackage{xcolor}
\usepackage{qcircuit}
\usepackage{braket}
\usepackage{float}
\usepackage{algorithm}
\usepackage{algorithmicx}
\usepackage{algpseudocode}
\usepackage{mathtools}
\usepackage{dirtytalk}
\usepackage{hyperref}
\usepackage{amsmath}
\usepackage{amsfonts}
\usepackage{amssymb}
\usepackage{amsthm}
\usepackage{newlfont}
\usepackage{array}
\usepackage{cite}

\usepackage{setspace}

\algnewcommand{\algorithmicand}{\textbf{ and }}
\algnewcommand{\algorithmicor}{\textbf{ or }}
\algnewcommand{\OR}{\algorithmicor}
\algnewcommand{\AND}{\algorithmicand}

\usepackage{authblk}

\providecommand{\keywords}[1]
{
  \textbf{\textit{Keywords---}} #1
}

\title{Scaling Limits of Quantum Repeater Networks}
\author[1]{Mahdi Chehimi}
\author[2]{Shahrooz Pouryousef}
\author[2]{Nitish K. Panigrahy}
\author[2]{Don Towsley}
\author[1]{Walid Saad}
\affil[1]{Wireless@VT, Bradley Department of Electrical and Computer Engineering, Virginia Tech, Arlington, VA USA}
\affil[2]{University of Massachusetts Amherst, Amherst, MA USA}
\affil[ ]{\textit{\{mahdic,walids\}@vt.edu}, \textit{\{shahrooz,nitish,towsley\}@cs.umass.edu}}

\begin{document}

\newcommand{\SP}[1]{\textcolor{blue}{Sharooz: #1}}
\newcommand{\MC}[1]{\textcolor{purple}{Mahdi: #1}}
\newcommand{\NP}[1]{\textcolor{green}{Nitish: #1}}
\newcommand{\DT}[1]{\textcolor{blue}{Don: #1}}

\maketitle
\begin{abstract}
Quantum networks (QNs) are a promising platform for secure communications, enhanced sensing, and efficient distributed quantum computing. However, due to the fragile nature of quantum states, these networks face significant challenges in terms of scalability. In this paper, the scaling limits of quantum repeater networks (QRNs) are analyzed. The goal of this work is to maximize the overall length, or \emph{scalability} of QRNs such that long-distance quantum communications is achieved while application-specific quality-of-service (QoS) requirements are satisfied. In particular, a novel joint optimization framework that aims at maximizing QRN scalability, while satisfying QoS constraints on the end-to-end fidelity and rate is proposed. The proposed approach optimizes the number of QRN repeater nodes, their separation distance, and the number of distillation rounds to be performed at both link and end-to-end levels. Extensive simulations are conducted to analyze the tradeoffs between QRN scalability, rate, and fidelity under gate and measurement errors. The obtained results characterize the QRN scaling limits for a given QoS requirement. The proposed approach offers a promising solution and design guidelines for future QRN deployments.
\end{abstract}
%Particularly, we develop a genetic algorithm-based solution to solve the proposed optimization problem, and
%, enabling secure and efficient long-distance quantum communication

\IEEEpeerreviewmaketitle

\vspace{-0.2cm}
\section{Introduction}\vspace{-0.1cm}
Quantum networks (QNs) are an emerging technology that holds tremendous promise for enabling secure and efficient long-distance quantum communications, precise quantum sensing, and significantly faster distributed quantum computing. However, the fragility of quantum states and their sensitivity to environmental impacts, noise, and losses restrict the scalability of QNs \cite{chehimi2022physics}. To overcome this limitation, quantum repeaters were introduced in order to form quantum repeater networks (QRNs) that allow sharing quantum states over longer distances. These networks employ \emph{entanglement swapping} \cite{briegel1998quantum} and \emph{entanglement distillation} \cite{bennett1996purification} operations to extend the communication range and enhance the quality, or \emph{fidelity}, of the transferred quantum states, respectively.

However, imperfections associated with these operations lead to unavoidable losses, and noise that could degrade the fidelity of quantum states. This poses serious challenges for the \emph{scalability} of QRNs, often defined as the overall QRN length or the product of number of repeater nodes and their separation distances. In particular, the number of repeater nodes in a QRN chain is constrained to a limit beyond which the resulting losses and fidelity degradation become unacceptable.

Designing end-to-end QRN chains for long-distance quantum communications between distant quantum nodes necessitates a detailed understanding and analysis of several factors including the installation of quantum repeater nodes, the number of such nodes, the distance between each neighboring pair of quantum repeaters, when each repeater should perform entanglement swapping, how many entanglement distillation operations should be performed on different entangled states in the chain, and how to schedule these operations. While some of these questions were investigated separately in the literature \cite{victora2020purification,hu2021long,chehimi2021entanglement,dai2020optimal,da2023requirements}, no prior work has explored their joint consideration in designing QRNs.

For instance, the work in \cite{victora2020purification} proposed an approach to optimize the distribution of entanglement in complex QRN architectures with imperfections by considering the interplay between bandwidth, distillation protocol, and path-finding algorithms. However, the authors in \cite{victora2020purification} did not analyze the scalability of their considered QRNs. Moreover, the work in \cite{hu2021long} proposed a novel entanglement distillation technique suitable for long-distance direct quantum communications. However, the authors in \cite{hu2021long} did not consider a QRN scenario in which quantum repeaters are present and distillation operations need to be scheduled. Similarly, the work in \cite{chehimi2021entanglement} optimized the entanglement generation rate (EGR) in quantum switch networks that neither included quantum repeaters nor incorporated entanglement distillation. Moreover, in \cite{dai2020optimal}, the authors studied optimal entanglement distribution in various QRN scenarios. When considering homogeneous repeater chains, the work in \cite{dai2020optimal} studied the impact of the total distance, or chain length, on the achievable entanglement distribution rate. However, the results in \cite{dai2020optimal} did not consider entanglement distillation operations. Finally, the study presented in \cite{da2023requirements} focused on a linear QRN, in which a small number of repeater nodes were strategically positioned between two end nodes that were 900 km apart. However, the authors in \cite{da2023requirements} did not address QRN scalability nor incorporated entanglement distillation operations. This shattered nature of state-of-the-art works on QRNs, their limitations, and the lack of a comprehensive analysis of QRN scalability represent a major setback for QRN development. This motivated us to thoroughly investigate QRN scalability issue, analyze its reliance on quality-of-service (QoS) requirements, and explore the different factors that affect the scalability of a QRN to achieve long-distance quantum communications.

%\MC{Stephanie \cite{da2023requirements} considered a linear path of fiber optics between two end nodes separated by 900 km. There are 16 location in between the end nodes where repeaters and heralding stations can be placed. The optimize over the placement of the repeaters in those possible locations, and consider two cases: 1) optimize the number of repeaters, and 2) restrict number of repeaters to specific values. They have a maximum of 7 repeaters. They investigate minimal hardware requirements for connecting two end nodes which are separated by roughly 900 km of real-world optical fiber in a linear QRN. They consider that there are 16 exact locations where the equipment (repeaters and heralding stations) can be placed in the network. They adopt the swap asap model, and model the swap noise as depolarizing noise. However, they assume that the gates and measurements applied by the end nodes when executing the quantum application (e.g., QKD and VBQC) are noiseless and instantaneous.}

Each of the aforementioned works \cite{victora2020purification,hu2021long,chehimi2021entanglement,dai2020optimal,da2023requirements} focused on one aspect of a QRN, like distillation for high fidelity, end-to-end rate enhancement, and scalability in terms of coverage. However, none of the existing works jointly studied QRN designs that take into consideration their scalability, entanglement distillation scheduling, and rate maximization. Unlike prior works, in this paper, we propose to jointly analyze all the aforementioned characteristics of QRNs. 

The main contribution of this work is to develop a holistic optimization framework to characterize the scalability of linear QRNs under different application-level QoS constraints on end-to-end fidelity and EGR. Towards achieving this goal, we make the following contributions: 
\begin{itemize}
    \item We explore the scalaing limits of QRNs, their connection to different QRN parameters and constraints, and their integration with QRN's noisy gates and operations under various application-specific QoS requirements.
   % \item We perform the first thorough analysis that integrates rate, fidelity, and scalability of QRNs under various application-specific QoS requirements.
    \item We formulate a novel optimization framework for maximizing the scalability of linear QRNs. We optimize the number of QRN repeaters, their separation distance, and the required amount of distillation rounds to satisfy the QoS constraints.
    \item We perform extensive simulations to analyze the tradeoffs between different QRN parameters, and identify their impacts on scalability. The results of our simulations demonstrate that the proposed framework is effective in providing meaningful insights into the scalability of linear QRNs under different QoS requirements, noise, and losses.
    %and the indirect relations between them, like the relation between the separation distance between QRN nodes and number of end-to-end distillation rounds.
    %Simulation results validate achieving insightful QRN scalability results based on the proposed framework under various QoS requirements, noise, and losses.
\end{itemize}

% \begin{itemize}
%     \item We perform the first thorough analysis that integrates rate, fidelity, and scalability of QRNs.
%     \item \MC{We optimize the number of quantum repeaters in a linear chain QRN, along with the distance between repeaters (separation distance) while satisfying stringent QoS requirements on the end-to-end fidelity and EGR}.
%     \item We \MC{optimize the scheduling of} entanglement distillation operations at both link-level and end-to-end level while taking into consideration their associated \MC{losses and measurement imperfections}.
%     \item Simulation results prove that ...
% \end{itemize}

The rest of this paper is organized as follows. Section \ref{sec_prelim} begins with a brief overview of necessary preliminary principles needed to develop the system model. Next, Section \ref{sec_system_model} describes the proposed system model of the QRN. The proposed optimization problem and its solution are presented in Section \ref{sec_optimization}. Then, in Section \ref{sec_simulations}, we conduct extensive simulations and experiments and analyse the key results. Finally, conclusions are drawn in Section \ref{sec_conclusion}.

\vspace{-0.075in}
\section{Quantum Preliminaries}\label{sec_prelim}
In this section, we provide a concise introduction to essential quantum concepts required to develop the system model.

\vspace{-0.075in}
\subsection{Quantum States}
In a QRN, link-level entangled (LLE) states are first generated between neighboring quantum nodes. We assume that each LLE state is of the form $\rho = W\ket{\psi_{00}}\bra{\psi_{00}} + \frac{1-W}{4}\Pi$. This state is known as the \emph{Werner state}. The \emph{fidelity} of such a quantum state is given as: $F_{L} = \frac{3W+1}{4}$.

\vspace{-0.075in}
\subsection{Entanglement Swapping}
Consider a linear chain QRN of $N+1$ quantum nodes and $N$ links. The fidelity of the obtained end-to-end entangled (E2E) state after performing $N-1$ swap operations on corresponding LLE states is given as \cite{briegel1998quantum}:
\begin{equation}\label{eq_nested_swaps_unequal}
\small
\begin{split}
    F_{E} = \frac{1}{4} + \frac{3}{4}&{\biggl(\frac{P_2(4\eta^2-1)}{3}\biggl)}^{N-1}\times\biggl(\frac{4F_{L,1}-1}{3}\biggl)\\
    &\times\biggl(\frac{4F_{L,2}-1}{3}\biggl)...\biggl(\frac{4F_{L,N}-1}{3}\biggl),
\end{split}
\end{equation}
where $P_2$ represents two-qubit gate fidelity, $\eta$ represents measurement fidelity of the entanglement swapping operation, and $F_{L,i}$ is the fidelity of the LLE state across link $i\in \{1,2,...,N\}$.

If all LLE states have the same fidelity, i.e., $F_{L,1} = F_{L,2} = \cdots = F_{L,N} = F_{\text{L}}$, then the output fidelity after performing entanglement swaps over $N$ links will be:
\begin{equation}\label{eq_nested_swaps_equal}
    S(F_{\text{L}},N) = \frac{1}{4} + \frac{3}{4}{\biggl(\frac{P_2(4\eta^2-1)}{3}\biggl)}^{N-1} {\biggl(\frac{4F_{\text{L}}-1}{3}\biggl)}^{N}.
\end{equation}

\subsection{Entanglement Distillation}
Throughout this work, we adopt a well-known symmetric IBM entanglement distillation protocol \cite{bennett1996purification}. In this protocol, the fidelity of the resulting state after performing one round of distillation on two identical Werner states, each with fidelity $F_{\text{in}}$, will be given by:
% \begin{equation}\label{eq_fidelity_distillation_identical}
%     f(F_{in}) = \frac{F_{in}^2 + (\frac{1-F_{in}}{3})^2}{F_{in}^2 + 2F_{in}(\frac{1-F_{in}}{3}) + 5(\frac{1-F_{in}}{3})^2}.
% \end{equation}

% P_{succ} = \left(F_{{\text{link}},i}+\frac{(1-F_{{\text{link}},i})}{3}\right)^2 + \left(\frac{2(1-F_{{\text{link}},i})}{3}\right)^2

\begin{equation}\label{eq_distill_noisy}
    f(F_{\text{in}}) = \frac{A(F_{\text{in}})\times B(\eta) + C(F_{\text{in}})\times D(\eta) + E(P_2)}{H(F_{\text{in}})\times B(\eta) + C(F_{\text{in}})\times 4D(\eta) + 4E(P_2)}
\end{equation}

where 
\begin{subequations}
\begin{alignat}{2}
    A(F_{\text{in}}) &= F_{\text{in}}^2 + \left(\frac{1-F_{\text{in}}}{3}\right)^2,\\
    B(\eta) &= \eta^2 + (1-\eta)^2,\\
    C(F_{\text{in}}) &= F_{\text{in}}\left(\frac{1-F_{\text{in}}}{3}\right) + \left(\frac{1-F_{\text{in}}}{3}\right)^2,\\
    D(\eta) &= 2\eta(1-\eta),\\
    E(P_2) &= \frac{1-P_2^2}{8P_2^2},\\
    H(F_{\text{in}}) &= F_{\text{in}}^2 + \frac{2}{3}F_{\text{in}}(1-F_{\text{in}}) + \frac{5}{9}(1-F_{\text{in}})^2.
\end{alignat}    
\end{subequations}

(\ref{eq_distill_noisy}) is subject to the constraint that $F_{\text{in}}\geq0.5$. The \emph{probability of success} for this entanglement distillation operation is given by: 
\begin{equation}
    P_{\text{S}}(F_{\text{in}}) = {P_2}^2 \times\bigl[ H(F_{\text{in}}) B(\eta) + C(F_{\text{in}}) 4D(\eta) + 4E(P_2)\bigr].
\end{equation}

We define a recursive function, $g(n, F_0)$, to calculate the final fidelity after performing $n$ rounds of entanglement distillation. Here, $F_0$ is the initial fidelity of the entangled quantum states that need to be distilled.
\begin{equation}\label{eq_recursive_function_g}
    g(n, F_0) = \begin{cases}
F_0 & \text{if } n = 0 \\
g(n-1, f(F_0)) & \text{otherwise}
\end{cases}.
\end{equation}

The fidelity after performing $i$ rounds of entanglement distillation, $F_i$, can be written in terms of the recursive function $g(n, F_0)$ as follows:
\begin{equation}
F_i = g(i, F_0) = g(i-1, f(F_0)) = f(F_{i-1}),
\end{equation}
Note that it is difficult to obtain a closed form expression for function $g(n,F_0)$ in terms of the initial fidelity and number of distillation rounds. As such, one has to resort to numerical computations to evaluate such functions.

%where $F_0$ is the initial fidelity of the entangled quantum states. We can repeat this recursive behavior for $F_{i-1}$, and so on, until we obtain an expression in terms of $F_0$. However, we notice that the expressions grow exponentially as the value of $i$ increases, which makes it extremely difficult to obtain a general closed-form expression for the recursive function of entanglement distillation, $g(n,F_0)$, in terms of the initial fidelity and the number of distillation rounds. 

% \begin{equation}
% F_i = \frac{(F_{i-1})^2 + \left(\frac{1-F_{i-1}}{3}\right)^2}{(F_{i-1})^2 + 2F_{i-1}\left(\frac{1-F_{i-1}}{3}\right) + 5\left(\frac{1-F_{i-1}}{3}\right)^2} = \frac{(F_{i-2})^4 + \left(\frac{1-F_{i-2}}{3}\right)^4}{(F_{i-2})^4 + 2F_{i-2}\left(\frac{1-F_{i-2}}{3}\right)^3 + 5\left(\frac{1-F_{i-2}}{3}\right)^4}.
% \end{equation}

%\vspace{-0.1in}
\section{System Model}\label{sec_system_model}
% \vspace{-0.05in}

\begin{figure}[t!]
\begin{center}%\vspace{-0.25in}
\centering
\includegraphics[width=\columnwidth]{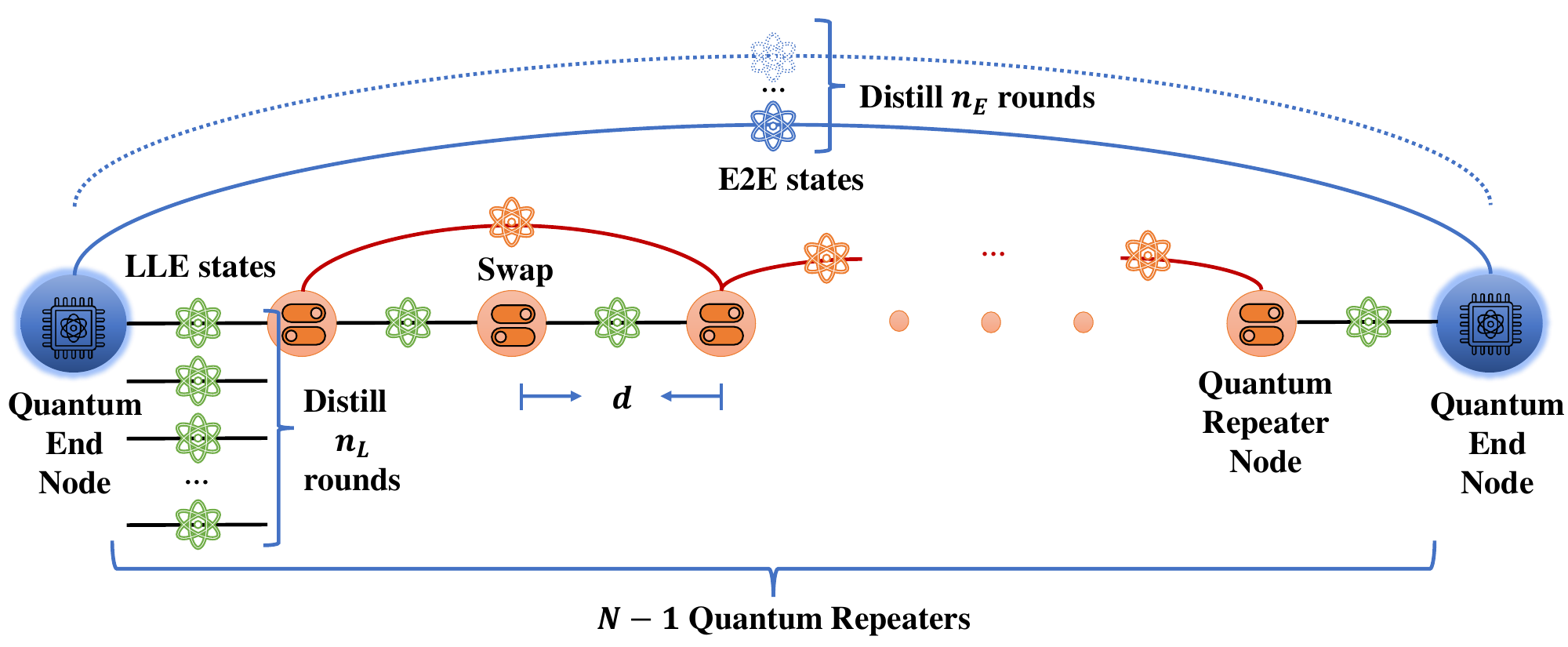}\vspace{-0.1in}
\caption{The considered linear QRN model.}
\label{fig_system_model}
\end{center}\vspace{-0.2in}
\end{figure}

We consider a linear QRN comprised of two end nodes connected through $N$ quantum links with $N-1$ quantum repeater nodes in-between. The QRN's end-goal is to create E2E states between the end quantum nodes with specified QoS requirements of \emph{EGR} and \emph{fidelity}, while extending the scale (or length) of the repeater chain as far as possible. Quantum repeaters are considered to be uniformly spaced, consistent with state-of-the-art literature \cite{da2023requirements}, and the separation distance between all neighboring pairs of quantum nodes is denoted as $d$. Moreover, we assume that all QRN nodes experience similar amounts of noise. The LLE states undergo several rounds of entanglement swapping and distillation before generating E2E states. Since each one of those operations affects the end-to-end fidelity and rate, they must be scheduled in a way that the QoS requirements are satisfied while maximizing scalability or length of the QRN. In particular, we define scalability of a linear QRN as the maximum length of the QRN, i.e., $N\times d$, for a given rate and fidelity constraint.

Based on state-of-the-art developments in QRN designs, we perform entanglement distillation only at the link and end-to-end levels, with no intermediate multi-hop operations \cite{nagayama2021towards}. Moreover, we assume that entanglement swapping operations are deterministic. Gates and measurements (whose fidelities are modeled as $P_2$ and $\eta$ in (\ref{eq_nested_swaps_unequal}), respectively \cite{briegel1998quantum}) are considered to be noisy in the QRN.

%\vspace{-0.2in}
\subsection{Link-level Fidelity and Rate}
The LLE states, which are Werner states shared between neighboring QRN nodes, are considered to be homogeneous. This means that they are generated at the link level with the same initial fidelity, $F_0$, and at the same rate, $R_0$, across all links in the QRN. Moreover, every pair of neighboring nodes will perform homogeneous entanglement distillation on the LLE states, i.e., the same number of distillation rounds are performed on each link. This results in the same output fidelity for LLE states, which depends on the initial fidelity $F_0$ and the \emph{link-level entanglement distillation parameter}, $n_{\text{L}}$. Here, $n_{\text{L}}$ represents the number of rounds of entanglement distillation performed at the link-level for each link. If we define $F_{{\text{L}},i}$ to be the fidelity before performing the $i$th distillation round at the link level, then the initial fidelity of the generated LLE state is represented as $F_0 = F_{{\text{L}},1}$. Accordingly, $F_{{\text{L}},i} = g(i-1, F_0), \quad \forall i\in\{1,2,...,n_{\text{L}}\}$, and the final fidelity of the link-level quantum states after performing $n_{\text{L}}$ entanglement distillation rounds will be: 
\begin{equation}
    F_{{\text{L}},n_{\text{L}}+1}(n_{\text{L}}, F_0) = f(F_{{\text{L}},n_{\text{L}}}) = g(n_{\text{L}}, F_0).
\end{equation}

%Denote the rate the initial LLE generation rate (before link-level distillation) are the same for all links, denoted as $R_0$. When those states are shared between neighboring quantum nodes, they need to trabel a distance $d$ over a quantum channel, e.g., optical fiber. 
Due to losses stemming from the interaction of the quantum states with the optical fiber, the probability of successfully creating an LLE state is $e^{-d/L_0}$, where $L_0$ is the attenuation length of the fiber optic, and $d$ is the distance between nodes \cite{rozpkedek2019near}. The initial EGR of an LLE state (before distillation) is $R_0e^{-d/L_0}$, where $R_0$ is the entanglement generating source repetition rate.

The LLE states are then subjected to the $n_{\text{L}}$ rounds of entanglement distillation, which result in reducing the LLE state generation rate. The resulting rate, $R_{\text{L}}$, after performing $n_{\text{L}}$ entanglement distillation rounds is found by accounting for the success probability of the entanglement distillation operations, the fidelity of each distillation round, and the exponential losses encountered during the establishment of the LLE states, as follows: \begin{equation}
    R_{\text{L}}(R_0, n_{\text{L}}, d, F_0) = \frac{R_0e^{-\frac{d}{L_0}}}{\prod_{i=1}^{n_{\text{L}}} \frac{2}{P_{\text{S}}(F_{{\text{L}},i})}}.
\end{equation}
\subsection{End-to-End Fidelity and Rate}
After performing $n_{\text{L}}$ link-level entanglement distillation operations, $N$ entanglement swap operations are performed on the distilled LLE states. Since we assume deterministic entanglement swap operations, swap operations only affect E2E state fidelity, without affecting the rate. Hence, the initial E2E state generation rate is $R_{\text{L}}(R_0, n_{\text{L}}, d, F_0)$. However, the initial E2E state fidelity before performing any end-to-end distillation operations, $F_{{\text{E}},1}$, is:
\begin{equation}
    F_{{\text{E}},1}(n_{\text{L}},F_0,N) = S(F_{\text{L},n_{\text{L}+1}}(n_{\text{L}}, F_0),N),
\end{equation}
where $S(\textbf{.})$ is given by (\ref{eq_nested_swaps_equal}). 

Next, we introduce the \emph{end-to-end entanglement distillation parameter}, $n_{\text{E}}$, which represents the number of entanglement distillation rounds performed at the end-to-end level. If we define $F_{{\text{E}},j}$ to be the fidelity before performing the $j$th distillation round at the end-to-end level, then, $F_{{\text{E}},j} = g(j-1, F_{{\text{E}},1}(n_{\text{L}},F_0,N)), \quad \forall j\in\{1,2,...,n_{\text{E}}\}$, and the final fidelity of the end-to-end level quantum states after performing $n_{\text{E}}$ entanglement distillation rounds is: 
\begin{equation}\label{eq_F_e2e}
    F_{{\text{E}},n_{\text{E}}+1}(n_{\text{E}}, n_{\text{L}}, F_0, N)) = g(n_{\text{E}}, F_{{\text{E}},1}(n_{\text{L}},F_0,N)).
\end{equation}

Moreover, the final end-to-end EGR, which accounts for the success probability of the $n_{\text{E}}$ end-to-end entanglement distillation operations is given as:
\begin{equation}\label{eq_R_e2e}
    R_{\text{E}}(n_{E},n_{L},N,d, R_0, F_0) = \frac{R_{\text{L}}(R_0, n_{\text{L}}, d, F_0)}{\prod_{j=1}^{n_{\text{E}}} \frac{2}{P_{\text{S}}(F_{{\text{E}},j})}}.
\end{equation}

% P_{succ} = \left(F_{{\text{e2e}},j}+\frac{(1-F_{{\text{e2e}},j})}{3}\right)^2 + \left(\frac{2(1-F_{{e2e},j})}{3}\right)^2

Th quantum application-specific QoS requirements impose minimum threshold constraints on the end-to-end fidelity and EGR in (\ref{eq_F_e2e}) and (\ref{eq_R_e2e}), respectively. Next, we formulate and solve a novel optimization problem for maximizing QRN scalability in the presence of such QoS constraints, while jointly finding the associated optimal scheduling of entanglement distillation operations. %This is the first work to jointly consider all of these factors while aiming at maximizing QRN scalability to achieve long-distance quantum communications.

\section{Optimization Formulation}\label{sec_optimization}
Now, we formulate the scalability problem in QRNs as an optimization problem whose goal is to maximize the length of a linear QRN while satisfying QoS constraints on end-to-end fidelity and rate. This problem explores how long (scalable) a linear QRN can be to achieve long-distance quantum communications while satisfying minimum QoS requirements. The available control variables are: 1) $n_{\text{L}}$, the number of link-level entanglement distillation operations, 2) $n_{\text{E}}$, the number of end-to-end entanglement distillation operations, 3) $N$, the number of intermediate entangled quantum states (this also corresponds to the number of repeating nodes, which is $N-1$), 4) $d$, the separation distance between neighboring nodes in the QRN. The QRN scalability optimization problem can be formulated as follows:
\begin{subequations}
\begin{alignat}{2}
\mathcal{P}1: \quad &\!\max_{N,d,n_{\text{L}},n_{\text{E}}}        &\quad& N\times d \label{eq:optProb}\\
&   s.t.               &      & R_{\text{E}}(n_{\text{E}},n_{\text{L}},N,d) \geq R_{\text{min}},\label{eq:constraint1}\\
&                  &      & F_{{\text{E}},n_{\text{E}}+1}(n_{\text{E}},n_{\text{L}},N) \geq F_{\text{min}},\label{eq:constraint2}\\
&                  &      & n_{\text{L}}\geq0, \quad\label{eq:constraint3}\\
&                  &      & n_{\text{E}}\geq0, \quad\label{eq:constraint4}\\
&                  &      & N\geq1, \quad\label{eq:constraint5}\\
&                  &      & d\geq 0, \quad\label{eq:constraint6}
\end{alignat}
\end{subequations}
where $R_{\text{min}}$ is the minimum required end-to-end EGR, and $F_{\text{min}}$ is the minimum required end-to-end fidelity. These two constraints capture the application-specific QoS requirements. %\MC{Moreover, the separation distance $d$ is not upper bounded ...}

%Moreover, the separation distance $d$, is considered to be upper bounded by a very large limit, $d_{\text{max}}$, since, in practical realizations of QRNs, the available quantum hardware has limitations on the number of attempts it can perform to generate entangled quantum states. Accordingly, and since increasing the separation distance between quantum nodes corresponds to transmitting the entangled quantum states over longer distances, i.e., more losses, the separation distance, theoretically, has an infinitely large upper bound. This is in order to satisfy the minimum rate QoS requirement.

We observe that the proposed optimization problem $\mathcal{P}1$ is a mixed-integer nonlinear programming problem. In order to solve $\mathcal{P}1$, and since the derivatives of the functions in (\ref{eq:constraint1}) and (\ref{eq:constraint2}) are not easily computed, it is typical to consider a derivative-free metaheuristic solution. However, in the conducted experiments, we notice that the effective ranges of the optimization control variables in the considered linear QRN can be explored using an exhaustive search algorithm to obtain the optimal solution of the optimization problem. Thus, we conduct most of our experiments using the exhaustive search algorithm. We also develop a metaheuristic solution based on the genetic algorithm (GA), and compare the performance with the exhaustive search algorithm. Such a metaheuristic solution is expected to play a vital role in future extensions of this work, where we consider heterogeneous and more complex QRN structures. %\MC{maybe we can say that we considered limited ranges of some parameters in order to be able to run the exhaustive search to get optimal solutions and to explore the impact of each parameter on the overall performance and the effective ranges and setp sizes that have an impactful effect.}

% Check 11c, if not satisfied, return -1
% if satisfied,
% calculate d by solving Re2e = Rmin
% if the obtained value is less or equal to 0, return -1,
% if obtained value of d is positive, 

%Final fidelity is not convex. If it is not possible to compute the gradient to apply gradient-based optimization. People did not attempt to do that.

% \subsection{No Rate Constraint}

% Use GA to find N, $n_link$, and $n_e2e$, then use this solution to 

%Paper on using genetic algorithm:

%https://iopscience.iop.org/article/10.1088/2058-9565/abfc93/pdf

\section{Simulation Results and Analysis}\label{sec_simulations}
In this section, we conduct extensive experiments to thoroughly investigate the relations between scalability, rate, and fidelity. Moreover, we analyze the impact of every QRN parameter on achievable end-to-end EGR and fidelity, in addition to the corresponding maximum QRN length. Throughout the experiments, we explore the effective ranges of all possible values for the different parameters in order to build the search space for the exhaustive search algorithm\footnote{The unbounded parameters are constrained by sufficiently large upper bounds to make the simulations tractable.}. Moreover, throughout the experiments, we fix $L_0 = 0.542$ km, which is the attenuation length used for the transmission of entangled states through optical fiber at the visible light wavelength \cite{hensen2015loophole}. Furthermore, unless stated otherwise, we consider the following \emph{default} quantum communication setup, where the noise parameters are set to $\eta = P_2 = 0.99$ \cite{harty2014high}, initial LLE state fidelity is $F_0 = 0.99$ \cite{harty2014high}, initial attempt EGR\footnote{Note that we do not consider multiplexing in the initial EGR attempts, which can be easily integrated into our framework to have higher numbers.} is $R_0 = 10^5$, a minimum end-to-end EGR requirement of $R_{\text{min}} = 1$ \cite{da2023requirements}, and a minimum end-to-end fidelity requirement $F_{\text{min}} = 0.5$ (which is the minimum fidelity needed to be able to perform entanglement distillation) \cite{bennett1996purification}.

\subsection{Impacts of Varying QoS Requirements}
\begin{figure}[t!]
\begin{center}%\vspace{-0.25in}
\centerline{\includegraphics[width=0.8\columnwidth]{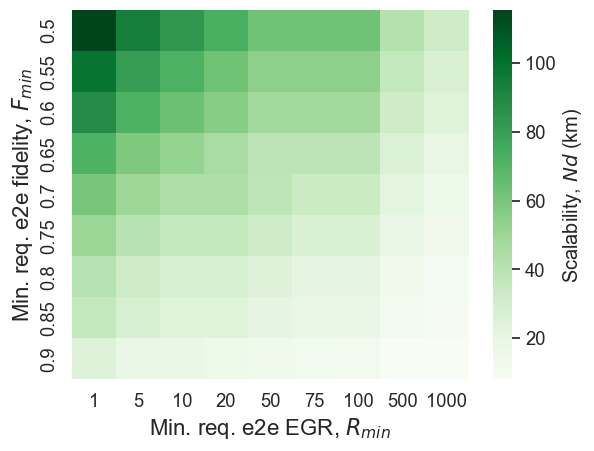}}\vspace{-0.15in}
\caption{Maximum QRN scalability vs $F_{\text{min}}$ and $R_{\text{min}}$.}
\label{fig_scalability_vs_QoS_new}
\end{center}\vspace{-0.35in}
\end{figure}

\begin{figure}[t!]
\begin{center}%\vspace{-0.25in}
\centerline{\includegraphics[width=0.7\columnwidth]{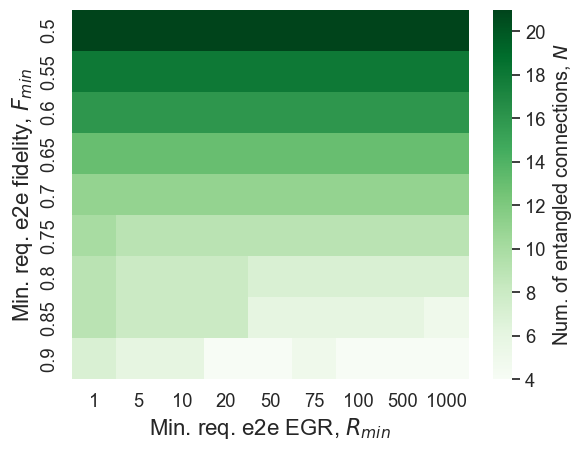}}\vspace{-0.15in}
\caption{Optimal number of entangled links, $N$, vs $F_{\text{min}}$ and $R_{\text{min}}$.}
\label{fig_N_vs_QoS_new}
\end{center}\vspace{-0.3in}
\end{figure}

First, we examine the effects of varying application-specific QoS requirements on scalability and number of QRN nodes. In Figure \ref{fig_scalability_vs_QoS_new}, we demonstrate how the optimal scalability varies as both $R_{\text{min}}$ and $F_{\text{min}}$ vary. Under best conditions, i.e., when $R_{\text{min}} = 1$ and $F_{\text{min}} = 0.5,$ we achieve a scalability of $115$ km. Similarly, Figure \ref{fig_N_vs_QoS_new} shows how the number of entangled links, $N$, varies with $R_{\text{min}}$ and $F_{\text{min}}$. From Figure \ref{fig_scalability_vs_QoS_new}, we observe that scalability decreases as the QoS requirements become more stringent. Furthermore, interestingly, Figure \ref{fig_N_vs_QoS_new} demonstrates that increasing $F_{\text{min}}$ leads to a more substantial decrease in $N$ compared to increasing $R_{\text{min}}$, which has only a minor effect on $N$. This implies that the rate requirement mainly influences the optimal $d$, which we will investigate next.

%the QRN controller tries to maximize the separation distance between the quantum nodes, which leads to restricting the number of repeater nodes in order to reduce the noise from increase in the travelled distance by the quantum states and the noise resulting from the swap operations. To further investigate this result, in Section \ref{sec_max_d}, we consider the case where we impose an upper bound on the separating distance, $d$, between quantum nodes in the QRN.  

% Similarly, we show in \ref{fig_d_vs_QoS} how the separation distance $d$ varies as the QoS requirements vary.

% \begin{figure}[t!]
% \begin{center}%\vspace{-0.25in}
% \centerline{\includegraphics[width=0.7\columnwidth]{img/d_vs_Rmin_and_Fmin.png}}\vspace{-0.3cm}
% \caption{Optimal separation distance, $d$, vs $F_{\text{min}}$ and $R_{\text{min}}$.}
% \label{fig_d_vs_QoS}
% \end{center}\vspace{-0.8cm}
% \end{figure}

\begin{figure}[t!]
\begin{center}\vspace{-0.05in}
\centerline{\includegraphics[width=0.7\columnwidth]{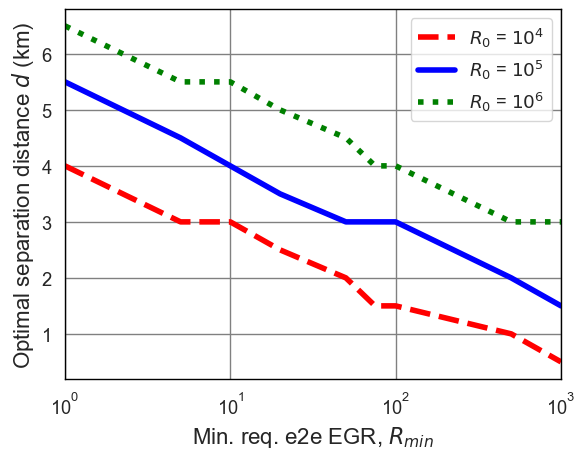}}\vspace{-0.15in}
\caption{Optimal separation distance, $d$, vs $R_{\text{min}}$ for different values of $R_0$.}
\label{fig_d_vs_Rmin_R_0}
\end{center}\vspace{-0.4in}
\end{figure}

The relationship between the optimal $d$, $R_{\text{min}}$, and $R_0$ is illustrated in Figure \ref{fig_d_vs_Rmin_R_0}. It can be observed that an increase in $R_{\text{min}}$ leads to a decrease in $d$. Similarly, decreasing the initial EGR attempts rate, $R_0$, also results in a decrease in $d$. We observe that even though there is no theoretical upper bound for $d$, its optimal values are not infinitely large. Figure \ref{fig_d_vs_Rmin_R_0} also indicates that achieving maximum scalability does not necessarily entail obtaining the largest possible separation distance $d$. This is due to the fact that increasing $d$ leads to entangled states traveling greater distances over optical fibers, which results in higher EGR losses. %Furthermore, increasing $d$ has an indirect impact on the maximum number of repeater nodes that can be incorporated in a QRN. \DT{Not always true if gate noise is large: Particularly, increasing $d$ leads to reducing the end-to-end rate, which constrains the number of end-to-end distillation rounds $n_{\text{E}}$ that could be performed by the repeater nodes in order to satisfy the minimum required end-to-end fidelity constraint.}\MC{we just want to point out that $d$ affects $n_L$, not necessarily decrease or increase it. Should we remove this whole sentence?} %We next consider the scalability performance when no link-level distillation is performed and only end-to-end distillation is available.  

% \subsection{Impact of Imposing Upper Limit Constraint on Separation Distance}\label{sec_max_d}
% Now, we consider the case where there is an upper limit, $d_{\text{max}}$, on the separation distance between QRN nodes, $d$. In particular, in Figure \ref{fig_constraint_dmax}, we show how the    

% \begin{figure}[t!]
% \begin{center}%\vspace{-0.25in}
% \centerline{\includegraphics[width=0.7\columnwidth]{img/Scalability_constraint_dmax.png}}%\vspace{-0.45in}
% \caption{Optimal QRN scalability vs $P_2$ and $\eta$ when $F_0 = 0.99$, $F_{\text{min}} = 0.5$, and $R_{\text{min}} = 1$.}
% \label{fig_constraint_dmax}
% \end{center}\vspace{-0.3in}
% \end{figure}

\subsection{Impact of having No Link-level Distillation}
\begin{figure}[t!]
\begin{center}%\vspace{-0.25in}
\centerline{\includegraphics[width=0.7\columnwidth]{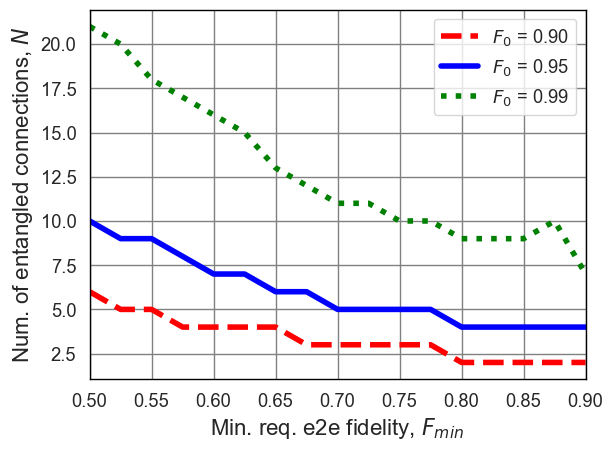}}\vspace{-0.15in}
\caption{Optimal $N$ vs $F_{\text{min}}$ when $n_{\text{L}} = 0$.}
\label{fig_N_no_link_distil}
\end{center}\vspace{-0.3in}
\end{figure}

\begin{figure}[t!]
\begin{center}\vspace{-0.05in}
\centerline{\includegraphics[width=0.7\columnwidth]{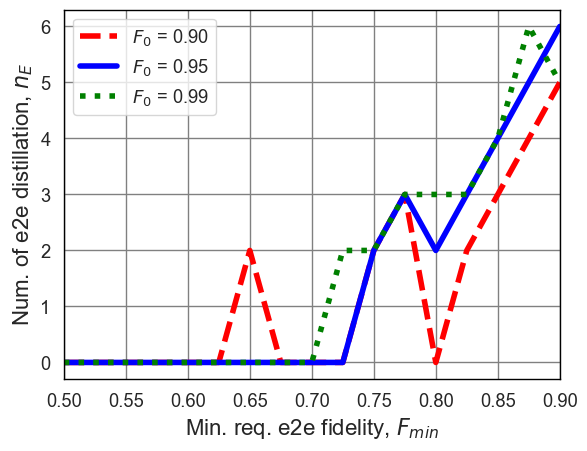}}\vspace{-0.15in}
\caption{Optimal $n_{\text{E}}$ vs $F_{\text{min}}$ when $n_{\text{L}} = 0$.}
\label{fig_n_e2e_no_link_distil}
\end{center}\vspace{-0.3in}
\end{figure}

\begin{figure}[t!]
\begin{center}\vspace{-0.05in}
\centerline{\includegraphics[width=0.7\columnwidth]{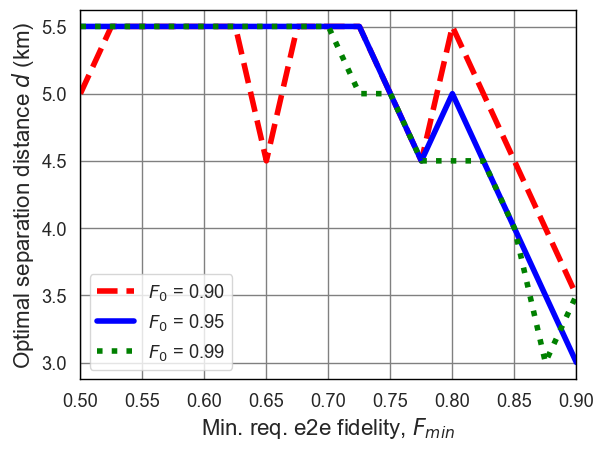}}\vspace{-0.15in}
\caption{Optimal separation distance, $d$, vs $F_{\text{min}}$ when $n_{\text{L}} = 0$.}
\label{fig_d_no_link_distil}
\end{center}\vspace{-0.4in}
\end{figure}

In the scenario where entanglement distillation is limited to end-to-end states only, i.e., $n_{\text{L}}=0$, we investigate its effect on the maximum scalability. To this end, we present in Figure \ref{fig_N_no_link_distil} the optimal $N$ as we vary $F_{\text{min}}$. Additionally, in Figures \ref{fig_n_e2e_no_link_distil} and \ref{fig_d_no_link_distil}, we analyze the variation of optimal $n_E$ and $d$ with $F_{\text{min}}$ when $n_L=0$.
%\ref{fig_n_e2e_no_link_distil}, and \ref{fig_d_no_link_distil}, 

By looking at both Figures \ref{fig_N_no_link_distil} and \ref{fig_d_no_link_distil}, we observe that the absence of link-level  distillation results in a notable decrease in scalability when the initial LLE state fidelity is decreased. This is because the link-level distillation operations are responsible for maximizing the LLE state fidelity before performing swap operations. When $n_{\text{L}}=0$, the number of swap operations, and, correspondingly, the number of repeater nodes in the QRN is significantly reduced. This is evident from Figure \ref{fig_N_no_link_distil} where we observe that $N$ drops by more than $50\%$ when the initial fidelity $F_0$ is only reduced by $4\%$, in the default setup with $F_{\text{min}}=0.5$. Moreover, by keeping $n_{\text{E}}$ low (as shown in Figure \ref{fig_n_e2e_no_link_distil}), we can get a high separation distance $d$ (as shown in Figure \ref{fig_d_no_link_distil}). However, when more stringent minimum fidelity requirements $F_{\text{min}}$ are needed, performing only end-to-end distillation may not be enough, and it may be necessary to also decrease the number of repeaters to minimize infidelities. 
%For instance, when $F_{\text{min}} = 0.75$, we observe from Figures \ref{fig_n_e2e_no_link_distil} and \ref{fig_d_no_link_distil} that when optimal $n_{\text{E}}$ is increased, the separation distance $d$ is decreased, and vice versa. This is because increasing the number of end-to-end entanglement distillation rounds reduces the achievable end-to-end rate, which limits the distance that entangled states can travel, i.e., the separation distance, which must be reduced (e.g., see the case of $F_{\text{min}} = 0.75$ in Figures \ref{fig_n_e2e_no_link_distil} and \ref{fig_d_no_link_distil}). This further clarifies the indirect relation between the separation distance, the number of repeaters, and the number of distillation rounds performed.

\subsection{Impact of having No End-to-end Distillation}
\begin{figure}[t!]
\begin{center}%\vspace{-0.25in}
\centerline{\includegraphics[width=0.7\columnwidth]{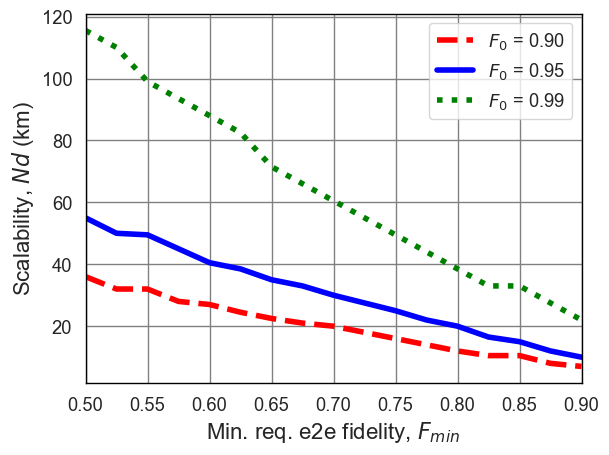}}\vspace{-0.15in}
\caption{Optimal scalability vs $F_{\text{min}}$ when $n_{\text{E}} = 0$.}
\label{fig_scalability_no_e2e_distil}
\end{center}\vspace{-0.35in}
\end{figure}

We now move on to examine another specific scenario where there are no end-to-end entanglement distillation operations, i.e., $n_{\text{E}} = 0$. In other words, we only perform entanglement distillation on LLE states. In Figure \ref{fig_scalability_no_e2e_distil}, we plot the optimal scalability as a function of the minimum end-to-end fidelity requirement $F_{\text{min}}$. Since end-to-end distillation operations are the ones responsible for correcting the noise and imperfections encountered during the swap operations, achieving high end-to-end fidelity (when $n_{\text{E}} = 0$) is significantly difficult.  Accordingly in Figure \ref{fig_scalability_no_e2e_distil}, we notice that, in order to meet the minimum fidelity requirements, the scalability must be compromised, and the number of repeater nodes must be considerably reduced.

%Next, we consider the special case in which no end-to-end level entanglement distillation operations are performed, i.e., $n_{\text{E}} = 0$. In this case, it is only possible to perform entanglement distillation on LLE states. To study the impact of this case on the optimal performance, we show, in Figure \ref{fig_scalability_no_e2e_distil}, the maximum scalability as a function of the minimum end-to-end fidelity requirement $F_{\text{min}}$. Since end-to-end distillation operations are the ones responsible for correcting the noise and imperfections encountered during the swap operations, achieving high end-to-end fidelity when $n_{\text{E}} = 0$ is significantly difficult. Accordingly, in Figure \ref{fig_scalability_no_e2e_distil}, we observe that in order to satisfy the minimum fidelity requirements, the scalabiliy will be sacrificed and the number of repeater nodes will be significantly reduced.

\subsection{Impact of Gate Noise and Imperfections}
\begin{figure}[t!]
\begin{center}%\vspace{-0.25in}
\centerline{\includegraphics[width=0.7\columnwidth]{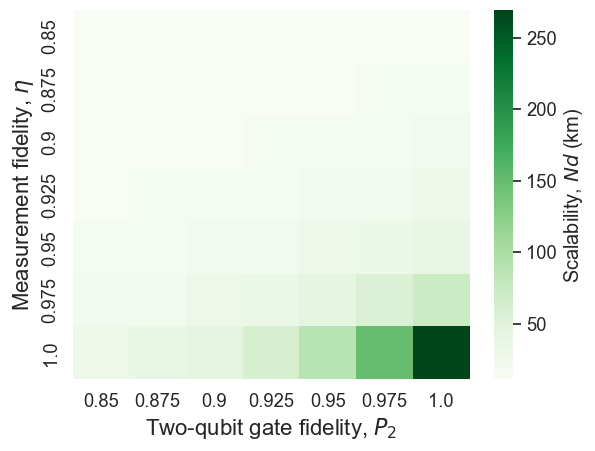}}\vspace{-0.15in}
\caption{Optimal QRN scalability vs $P_2$ and $\eta$ when $F_0 = 0.99$, $F_{\text{min}} = 0.5$, and $R_{\text{min}} = 1$.}
\label{fig_scalability_vs_noise}
\end{center}\vspace{-0.35in}
\end{figure}

Next, we study the impact of the noise introduced during distillation and swapping on the scalability of a QRN. Here, we consider the \emph{default} quantum communication setup parameters in order to focus on the direct impact of noise on scalability. From Figure \ref{fig_scalability_vs_noise}, we observe that maximum QRN scalability significantly varies as a function of noise. It is clear from Figure \ref{fig_scalability_vs_noise} that the infidelities associated with imperfect measurements (whose fidelity is represented by $\eta$) have a greater impact on QRN scalability compared to the impact of two-qubit gates noise (where the gate fidelity is represented by $P_2$). Additionally, we also observe from Figure \ref{fig_scalability_vs_noise} that, in the absence of device imperfections, the QRN can achieve a scalability of approximately 267 km. This value can be further improved by incorporating multiplexing techniques in the initial EGR attempts.
%for the case in which the QoS requirements are not very stringent, the different noise levels would not affect the optimal separation distance $d$ between neighboring quantum nodes, which was $5.51$ km in all cases, since no other parameters were varied.

%\MC{In practical setups, the fidelity of such operations (measurements and gates) do not usually fall below $95\%$, which is sufficient to have an acceptably scalable QRN performance. However, if the number of applied gates and performed measurmeents is significantly large, their corresponding noise accumulate and it becomes difficult to have a scalable QRN in such scenarios [cite].} 

\vspace{-0.05in}
\subsection{Performance of Metaheuristic GA Solution}
\begin{figure}[t!]
\begin{center}%\vspace{-0.25in}
\centerline{\includegraphics[width=0.7\columnwidth]{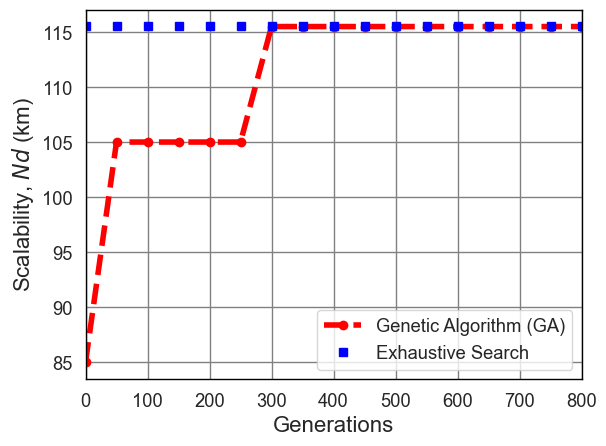}}\vspace{-0.15in}
\caption{Evolution of scalability in GA vs the number of generations, and compared to exhaustive search.}
\label{fig_GA_evolution}
\end{center}\vspace{-0.3in}
\end{figure}

Finally, we implement a genetic algorithm (GA) to solve the QRN scalability optimization problem. We then compare the scalability obtained through the GA with the scalability obtained through the exhaustive search algorithm as shown in Figure \ref{fig_GA_evolution}. Here, we adopt the \emph{default} quantum communication setup parameters defined earlier. In terms of solution optimality, the exhaustive search algorithm yielded a maximum scalability of 115.521 km. In comparison, the GA algorithm produced a scalability of 115.512 km, resulting in an error gap of less than $0.01\%$ between the two solutions. Additionally, the GA approach was able to converge to the best  solution within 300 generations, which is a $52\%$ reduction in running time compared to the exhaustive search algorithm. Although the running time of the exhaustive search algorithm for the homogeneous linear QRN scenario was reasonable for achieving optimality in our experiments, the proposed GA solution will be useful for future extensions that consider more complex and heterogeneous QRN scenarios.

%Moreover, in terms of running time, GA converged to the suboptimal solution after 300 generations, and it saved about $52\%$ of the running time that exhaustive search needed to find the optimal solution. As mentioned earlier, the running time of the exchaustive search algorithm for the studied homogeneous linear QRN scenario is affordable, so we adopted it in conducting the experiments in order to guarantee optimality. The proposed GA solution will be adopted in future extensions that include more complex and heterogeneous QRN scenarios. 
%which means that the suboptimal solution obtained using GA is very close to the optimal solution obtained using exhaustive search

%GA with discretized d vs GA with continuous d (but set constraint 11b to equality, and pick d that gives the minimum e2e rate)

\vspace{-0.05in}
\section{Conclusion}\label{sec_conclusion}
In this paper, we have taken a comprehensive approach to address the problem of scalability in homogeneous linear QRNs, while taking into consideration various QoS requirements. Specifically, we have proposed a new optimization framework that jointly maximizes QRN scalability and the entanglement distillation operations to meet the QoS requirements for both end-to-end fidelity and rate. Furthermore, we have conducted extensive experiments to investigate the impact of different QRN parameters on scalability and have identified the indirect relationships between variables such as the number of repeater nodes, their separation distance, and the number of distillation rounds on different QRN levels. Going further, we would like to analyze the scalability of more complex and non-homogeneous QRN structures.

\vspace{-0.07in}
\section*{Acknowledgment}\vspace{-0.05in}
This research was supported in part by the NSF grant CNS-1955744, NSF-ERC Center for Quantum Networks grant EEC-1941583, and MURI ARO Grant W911NF2110325.

\vspace{-0.05in}
\begin{spacing}{0.95}
\bibliographystyle{IEEEtran}
\bibliography{References}
\end{spacing}

% that's all folks
\end{document}